\documentclass[twocolumn,hyperpdf,
amsmath,amssymb,
aps,prd,10pt,
superscriptaddress,nofootinbib,preprintnumbers]{revtex4-1}

\usepackage{float}

\usepackage{amsfonts}
\usepackage{bm,graphicx}
\usepackage{hyperref}

\pdfoutput=1 
\usepackage{amssymb,amsmath,hyperref}
\usepackage{color}
\usepackage{graphicx}
\usepackage{bm}
\usepackage{amsmath,amssymb}
\usepackage{mathtools}
\usepackage{graphicx}
\usepackage{bm}
\usepackage{comment} 
\usepackage{subfigure}
\usepackage{array}
\usepackage{multirow}

\newcommand{\nn}{\nonumber}

\newcommand{\F}{F_{\pi\rho}}
\newcommand{\RLL}{\mathcal{R}}

\hypersetup{colorlinks, linkcolor = [rgb]{0,0.0,0.75}, citecolor = [rgb]{0,0.0,0.75}, urlcolor = [rgb]{0,0.0,0.75}}

\begin{document}

 \preprint{\vbox{
\hbox{JLAB-THY-15-2106} }}

 \preprint{\vbox{\hbox{DAMTP-2015-40} }}


\title{Resonant $\pi^+\gamma\to\pi^+\pi^0$ Amplitude from Quantum Chromodynamics}

\author{Ra\'ul A. Brice\~no}
\email[e-mail: ]{rbriceno@jlab.org}
\affiliation{Thomas Jefferson National Accelerator Facility, 12000 Jefferson Avenue, Newport News, VA 23606, USA}
\author{Jozef J. Dudek}
\affiliation{Thomas Jefferson National Accelerator Facility, 12000 Jefferson Avenue, Newport News, VA 23606, USA}
\affiliation{Department of Physics, Old Dominion University, Norfolk, VA 23529, USA}
\author{Robert G. Edwards}
\affiliation{Thomas Jefferson National Accelerator Facility, 12000 Jefferson Avenue, Newport News, VA 23606, USA}
\author{Christian J. Shultz}
\affiliation{Department of Physics, Old Dominion University, Norfolk, VA 23529, USA}
\author{Christopher~E.~Thomas}
\affiliation{Department of Applied Mathematics and Theoretical Physics, Centre for Mathematical Sciences, University of Cambridge, Wilberforce Road, Cambridge, CB3 0WA, UK}
\author{David J. Wilson}
\affiliation{Department of Physics, Old Dominion University, Norfolk, VA 23529, USA}

\collaboration{for the Hadron Spectrum Collaboration}

\begin{abstract}
We present the first \emph{ab initio} calculation of a radiative transition of a hadronic resonance within quantum chromodynamics (QCD). We compute the amplitude for $\pi\pi \to \pi\gamma^\star$, as a function of the energy of the $\pi\pi$ pair and the virtuality of the photon, in the kinematic regime where $\pi\pi$ couples strongly to the \emph{unstable} $\rho$ resonance. This exploratory calculation is performed using a lattice discretization of QCD with quark masses corresponding to $m_\pi \approx 400$~MeV. We obtain a description of the energy dependence of the transition amplitude, constrained at 48 kinematic points, that we can analytically continue to the $\rho$ pole and identify from its residue the $\rho \to \pi\gamma^\star$ form factor.

 \end{abstract}

\maketitle

\paragraph{Introduction:}~
The electromagnetic transitions of the nucleon into \emph{unstable resonant} $N^\star$ excitations are primary tools in the experimental study of nucleon structure and spectroscopy~\cite{Aznauryan:2012ba}. 
These processes give us insight into the mechanisms that lead to the formation of the low-lying and excited hadrons from the basic quark and gluon building blocks of quantum chromodynamics (QCD). It is crucial to have a complementary theoretical program that connects physically observed transitions to QCD. One major challenge in studying these transitions is their resonant nature, where the $N^\star$ excitation decays rapidly to asymptotic scattering states composed of two or more stable hadrons. To investigate these processes within QCD, one needs a nonperturbative framework that can accommodate resonant behavior, and presently, lattice QCD is the only available tool to evaluate such observables while making only controlled approximations. Its implementation for the determination of properties of hadron resonances is still at an exploratory stage, and in this work we will extend the exploration into a new area with the first calculation of a radiative production amplitude of an unstable hadronic resonance from QCD.

Before attempting the more complicated baryonic case of $\gamma^\star  N \to N^\star \to N\pi$, we will consider a simpler problem featuring only mesons, $\pi\gamma^\star \to \rho \to \pi\pi$, which in addition to serving as the first of a new class of observables to be studied, is itself of significant phenomenological interest. The amplitude for this process is related to the hadronic contribution to the anomalous magnetic moment of the muon~\cite{Colangelo:2014dfa, Colangelo:2014pva}, the chiral anomaly~\cite{Wess:1971yu, Witten:1983tw}, and the $\rho \to \pi \gamma$ radiative decay rate~\cite{Huston:1986wi, Capraro:1987rp}, and appears in meson-exchange models of nuclear structure~\cite{Marcucci:2015rca}. The $\rho \to \pi \gamma^\star$ transition has been previously studied using lattice methods (see, for example, Refs.~\cite{Shultz:2015pfa, Owen:2015gva, Owen:2015fra}), but prior to this work the $\rho$ has always been treated as a stable hadron, incapable of decay to $\pi\pi$, in contrast to how it appears in experiment. This approximation, which is uncontrolled for light quark masses such as those used in Refs.~\cite{Owen:2015gva, Owen:2015fra} is removed in the present work.

The perturbative nature of quantum electrodynamics ensures that to an excellent approximation the $\pi\pi \to \pi\gamma^\star$ amplitude can be obtained from matrix elements of the electromagnetic current, ${\mathcal{J}}^{\mu}= \frac{2}{3}\bar{u}\gamma^\mu u-\frac{1}{3}\bar{d}\gamma^\mu d$,
 \begin{align}
 \label{eq:infinite_volue_amp}
\mathcal{H}_{\pi\pi,\pi\gamma^\star}^{\mu}
=
{ \big\langle  \rm{out}; \pi,P_\pi \big| {\mathcal{J}}^{\mu}_{x=0} \big| \rm{in};\pi\pi,P_{\pi\pi},\ell=1 \big\rangle},
\end{align}
where the $\pi\pi$ state has been projected onto an $\ell=1$ partial wave, and where $P_{\pi\pi}$ and $P_{\pi}$ are the 4-momenta of the $\pi\pi$ and $\pi$ states, respectively. We will determine this amplitude as a function of the c.m. frame energy of the pion pair $E_{\pi\pi}^\star$ and the virtuality of the photon, $Q^2 = - (P_\pi - P_{\pi\pi})^2$, by evaluating correlation functions using lattice QCD. 

Lattice QCD calculations are performed in a finite, discretized Euclidean spacetime, and this introduces three length scales into the theory: the lattice spacing, $a$, and the spatial ($L$) and temporal ($T$) extents of the volume. For studies of stable hadrons not featuring heavy quarks, provided $m_\pi T,\,  m_\pi L \gg 1$ and $a \ll 1\,\mathrm{fm}$, the typical length scale associated with hadrons, these approximations introduce only small and controllable systematic errors.

The restriction to a finite volume in space prohibits the definition of asymptotic states, making the relationship between few-body observables obtained via lattice QCD and the scattering amplitudes of infinite-volume QCD somewhat nontrivial. As has been extensively explored in the literature, scattering amplitudes of two-body~\cite{Luscher:1986pf, Luscher:1990ux, Rummukainen:1995vs, Kim:2005gf, Christ:2005gi, Briceno:2012yi, Hansen:2012tf, Briceno:2014oea} and three-body systems~\cite{Polejaeva:2012ut,Briceno:2012rv,Hansen:2015zga,Hansen:2014eka} can be determined from the spectrum of eigenstates of QCD in a finite volume. Such spectra can be obtained from two-point correlation functions within lattice QCD, and the energydependence of hadron scattering amplitudes can be inferred -- by analytically continuing these amplitudes to complex values of the scattering energy, poles can be found, with the pole positions providing the mass and width of hadronic resonances. For an example see the recent determination of kaon resonant excitations in coupled-channel $\pi K, \eta K$ scattering~\cite{Dudek:2014qha, Wilson:2014cna}, and the $\rho$ resonance for lighter quark masses where $\pi\pi, K\overline{K}$ can be coupled~\cite{Wilson:2015dqa}.

The extension of the formalism to the case where an external (e.g. electroweak) current  causes a transition from a single stable hadron to a pair of hadrons was presented by Lellouch and L\"uscher. They demonstrated that one can constrain such an amplitude using hadronic matrix elements of the currents evaluated in a finite volume~\cite{Lellouch:2000pv}. Their work focused on the implications of this formalism for $K \to \pi\pi$ decays, where $\pi\pi$ is in an $S$ wave (see Refs.~\cite{Bai:2015nea,  Blum:2012uk, Boyle:2012ys, Blum:2011pu, Blum:2011ng} for numerical implementations), and this has been subsequently extended to other systems of interest~\cite{Lin:2001ek, Kim:2005gf, Christ:2005gi, Hansen:2012tf, Agadjanov:2014kha, Meyer:2011um,Meyer:2012wk, Feng:2014gba}.

Recently, these ideas were extended to accommodate more generic processes featuring an external current~\cite{Briceno:2015csa, Briceno:2014uqa}, and in this work it was shown that the transition amplitude, $\mathcal{H}_{\pi\pi,\pi\gamma^\star}^{\mu}$, can be obtained from finite volume matrix elements of the vector current,
 \begin{align}
 \label{eq:Amp_to_matelem}
\frac{|{\mathcal{H}_{\pi\pi,\pi\gamma^\star}^{\mu} }|}{L^3}
\sqrt{\frac{\RLL}{2E_\pi}}=
\Big|{   \prescript{}{L}{\big\langle} \pi; P_\pi, \Lambda_{\pi} \big|{\mathcal{J}}^{\mu}_{x=0} \big| \pi\pi; P_{\pi\pi}, \Lambda_{\pi\pi} \big\rangle}_L \Big|,
\end{align}
where ${\RLL}$ is the residue of the finite-volume two-hadron propagator, which depends on the $\pi\pi$ 4-momentum, the cubic irreducible representation ($\Lambda_{\pi\pi}$), the lattice volume ($L\times L \times L$), and the $\pi\pi$ elastic scattering amplitude. The hadronic finite-volume eigenstates carry labels, $\Lambda$, which indicate in which irreducible representation, or ``irrep'', of the reduced rotational symmetry of the cubic lattice they lie. We point the reader to Ref.~\cite{Briceno:2014uqa} for a detailed derivation and definition of $\RLL$. Equation~(\ref{eq:Amp_to_matelem}) is an approximation of the result presented in Ref.~\cite{Briceno:2014uqa} -- we have ignored contributions due to mixing with higher partial waves, which are both kinematically and dynamically suppressed in the energy regime of interest.~\footnote{In Ref.~\cite{Dudek:2012xn} it was demonstrated that the $\ell \geq3$ $\pi\pi$ scattering phase shifts are consistent with zero. 
This assures one that the only resonance present that couples to the $I=1$ $\pi\pi$ channel in this kinematic regime is the $\rho$-meson and so the $\ell=1$ transition amplitude is the dominant contribution.}


\paragraph{Lattice QCD calculation:} 
We use an anisotropic Symanzik improved gauge and Clover fermion actions with $N_f=2+1$ dynamical fermions. The quark masses are chosen so that $m_\pi \sim 400$~MeV~\cite{Lin:2008pr}, and we use a spacetime volume of $(L/a_s)^3\times (T/a_t)=20^3\times 128$, where $a_s$ and $a_t$ are the spatial and temporal lattice spacings with $a_s/a_t =3.444(6)$ and $a_s\approx 0.12$~fm. In Ref.~\cite{Dudek:2012gj} it was demonstrated that exponential corrections associated with the finite volume of this lattice lead to subpercent corrections. 

We construct three-point correlation functions using the technology presented in Ref.~\cite{Shultz:2015pfa}. We use variationally optimized $\pi$ and isospin-1 $\pi\pi$ operators, $\Omega^{[\Lambda_\pi]}_{\pi}$ and $\Omega^{[\Lambda_{\pi\pi}]}_{\pi\pi}$, respectively, that have been subduced to the desired irrep, $\Lambda$, of the appropriate little group of the octahedral group~\cite{Thomas:2011rh}. These operators have been previously obtained in the determination of the spectrum from two-point correlation functions~\cite{Dudek:2012xn}. Inserting the vector current we have three-point functions,
\begin{align}
\label{eq:3pointfunc}
\big\langle 0 \big|
\Omega^{[\Lambda_\pi]}_{\pi}(\Delta t,\,  \textbf{P}_{\pi}) \,
{\mathcal{J}}^{\mu}(t,\textbf{P}_{\pi}\!-\!\textbf{P}_{\pi\pi})\, 
\Omega^{[\Lambda_{\pi\pi}]\dag}_{\pi\pi}(0, \textbf{P}_{\pi\pi})
\big|0\big\rangle,
\end{align}
and we will present results extracted from correlation functions computed with Euclidean time separation, ${\Delta t= 32 a_t}$, excluding Wick contractions where the current couples to a disconnected quark loop~\footnote{These exactly vanish in the $SU(3)$ flavor limit and are expected to be suppressed for the quark masses used.}. 
${\mathcal{J}}^{\mu}$ is the tree level improved Euclidean vector current~\cite{Shultz:2015pfa}, which is renormalized by insisting the $\pi$ form factor be 1 at $Q^2=0~\rm{GeV}$, giving a multiplicative renormalization of $Z_V=0.833(9)$.
By inserting a complete set of finite volume QCD eigenstates in Eq.~(\ref{eq:3pointfunc}), and evolving the operators to the origin of Euclidean time, one can determine $ \prescript{}{L}{\big\langle} \pi; P_\pi, \Lambda_{\pi} \big|{\mathcal{J}}^{\mu}_{x=0} \big| \pi\pi; P_{\pi\pi}, \Lambda_{\pi\pi} \big\rangle_L$ from the time dependence of the correlation function~\cite{Shultz:2015pfa}. 

Consideration of various momenta, ${\textbf{P} = \frac{2\pi}{L}[n_x, n_y, n_z]}$, allowed by the periodic boundary conditions, leads to determination of the matrix element at 48 distinct kinematic points. Eight different discrete $E_{\pi\pi}^\star$ values feature, corresponding to the finite-volume eigenstates of $\pi\pi$ in various irreps, and discrete values of photon virtuality in the range $-0.4 \le (Q/{\rm GeV})^2 \le 1$ are sampled.


\vspace{0.5cm}
\paragraph{The transition amplitude and $\rho\to\pi\gamma^\star$ form factor:}
The $\pi\pi \to \pi\pi $ $P$-wave elastic scattering amplitude, expressed via a phase shift, $\delta_1(E_{\pi\pi}^\star)$, was determined from the lattice QCD finite-volume spectrum in Ref.~\cite{Dudek:2012xn}. With this in hand we may evaluate $\RLL$ in Eq.~(\ref{eq:Amp_to_matelem}) and determine the transition amplitude from the finite-volume matrix elements.

The transition amplitude is a Lorentz vector, and it has a kinematic decomposition,
\begin{align}
\label{eq:Apipipi}
\mathcal{H}_{\pi\pi,\pi\gamma^\star}^{\mu}=
\epsilon^{\mu\nu\alpha\beta}  \, P_{\pi,\nu}\, P_{\pi\pi,\alpha} \, \epsilon_{\beta}(\lambda_{\pi\pi},\textbf{P}_{\pi\pi})\, \tfrac{2}{m_\pi} \mathcal{A}_{\pi\pi,\pi\gamma^\star},
\end{align}
where $\mathcal{A}_{\pi\pi,\pi\gamma^\star}(E_{\pi\pi}^\star, Q^2)$ is a Lorentz scalar and $\epsilon_{\beta}$ is the polarization vector of the $P$-wave $\pi\pi$ state with $\lambda_{\pi\pi}$ being its helicity.  

In the energy region we consider, the transition amplitude will be sharply peaked due to the $\rho$ pole. Defining a $\rho\to\pi\gamma^\star$ form factor, $\F(E^\star_{\pi\pi},Q^2)$, we may write the amplitude,
\begin{align}
\label{eq:AtoF}
{\mathcal{A}}&_{\pi\pi,\pi\gamma^\star}(E^\star_{\pi\pi},Q^2)  \nonumber \\
  &= {\F(E^\star_{\pi\pi},Q^2)} \, \sqrt{\frac{8\pi }{q^\star_{\pi\pi}\, \Gamma_{1}(E^\star_{\pi\pi})}}\, \sin\delta_1(E_{\pi\pi}^\star) \, e^{i\delta_1(E_{\pi\pi}^\star)},
\end{align}
which is proportional to the elastic $\pi\pi \to \pi\pi$ scattering amplitude $\mathcal{M}_{\pi\pi}^{\ell=1}=\frac{8\pi E^\star_{\pi\pi}}{ q^\star_{\pi\pi}} \sin \delta_1 \, e^{i \delta_1}$ with $q^\star_{\pi\pi} = \frac{1}{2}\sqrt{ E^{\star 2}_{\pi\pi} - 4 m_\pi^2}$. In Refs.~\cite{Agadjanov:2014kha, Briceno:2015csa} it was demonstrated that this parametrization is consistent with the analyticity and unitarity constraints required in scattering theory. The presence of the energy-dependent $\rho\to\pi\pi$ strong decay width, $\Gamma_1$, can be understood in the context of an effective field theory where the rescattering of the final $\pi\pi$ states is mediated by a fully dressed $\rho$ resonance (see the appendices of Ref.~\cite{Briceno:2015csa}). 

The derivative of the phase shift, $\frac{d\delta_1}{dE^\star_{\pi\pi}}$, appears in $\RLL$ -- to compute it we use a sensible parametrization for $\delta_1(E^\star_{\pi\pi})$, the relativistic Breit-Wigner function~\cite{Dudek:2012xn}.  

\begin{figure}[t]
\begin{center}
\hspace{-.5cm}
\includegraphics[scale=0.38]{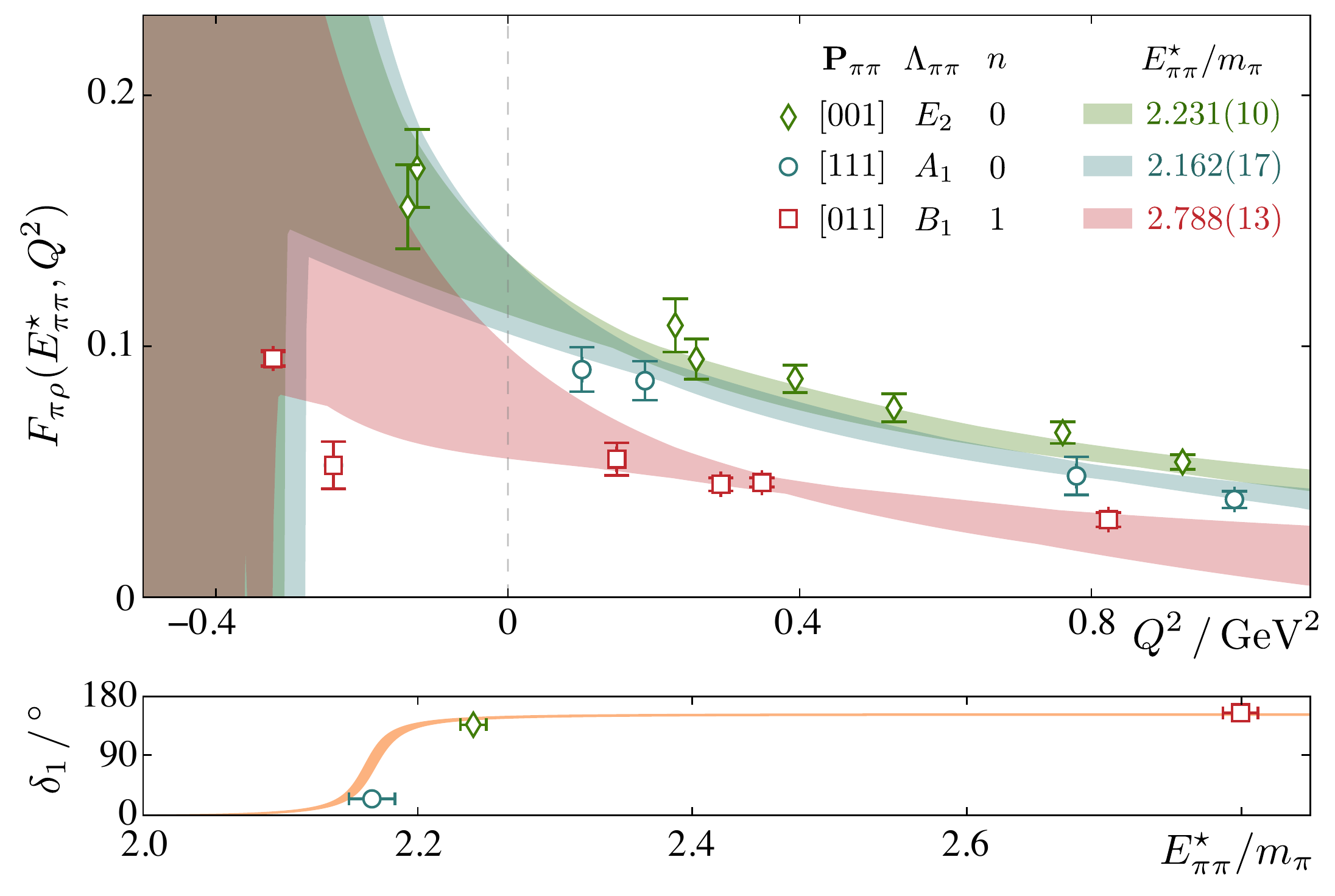}\vspace{-.5cm}
\caption{The points appearing in the upper panel depict the form factor determined from lattice QCD for three $\pi\pi$ energy levels. The index $n$ labels the order in which the state appears in the spectrum. Also shown are the fits of the form factor performed using Eq.~(\ref{eq:paramansatz}) and evaluated at the three $\pi\pi$ energies. The corresponding $P$-wave phase shift and $\pi\pi$ energy is shown on the lower panel.  }
\label{fig:Fpirho_BW_irrep_plots}
\vspace*{-.6cm}
\end{center}
\end{figure}

In Fig.~\ref{fig:Fpirho_BW_irrep_plots} we present the computed form factor for three of the eight $\pi\pi$ energies studied, using the Breit-Wigner parametrization for the phase shift, where we observe that both spacelike and timelike $Q^2$ kinematics are sampled. It is evident that $\F(E^\star_{\pi\pi},Q^2)$ has only a mild dependence on $E^\star_{\pi\pi}$, with the sharply peaked resonant behavior having been captured by the $\sin \delta_1(E^\star_{\pi\pi})$ factor in Eq.~(\ref{eq:AtoF}).

To analytically describe the $E^\star_{\pi\pi}$ and $Q^2$ dependence of the form factor we introduce an ansatz,
\begin{align}
h^{[\{\alpha,\beta\}]}(E^\star_{\pi\pi},Q^2)&=\nn\\
&\hspace{-1.0cm}\frac{\alpha_1}{1+\alpha_2Q^2+\beta_1(E^{\star2}_{\pi\pi}-m^2_0)}+\alpha_3Q^2+\alpha_4Q^4
\nn\\
&\hspace{-1.0cm}+\alpha_5 \exp\left[{-\alpha_6 Q^2-\beta_2(E^{\star2}_{\pi\pi}-m_0^2) }\right]\nn\\
&\hspace{-1.0cm}+\beta_3(E^{\star2}_{\pi\pi}-m_0^2)+\beta_4(E^{\star4}_{\pi\pi}-m^4_0),
\label{eq:paramansatz}
\end{align}
where the parameters $\alpha_i$ and $\beta_i$ are to be fitted and the constant $m_0$ is fixed to $2.1805\, m_\pi$ to coincide with the real part of the $\rho$ mass. To fit the form factor, we vary the form being used by setting a subset of these coefficients  to zero and thus consider over 15 different fit functions. We also consider fits where the points in the timelike $Q^2$ region are excluded. From all fits performed, we retain only those that have a $\chi^2/\rm DOF\leq 1.5$, and we find that no fit lacking $E^\star_{\pi\pi}$-dependent terms satisfies this. The bands shown in Figure~\ref{fig:Fpirho_BW_irrep_plots} reflect the parametrization variation as will the uncertainties on all quantities quoted below.

\begin{figure}[t]
\begin{center} 
\hspace{-.5cm}
\includegraphics[scale=0.37]{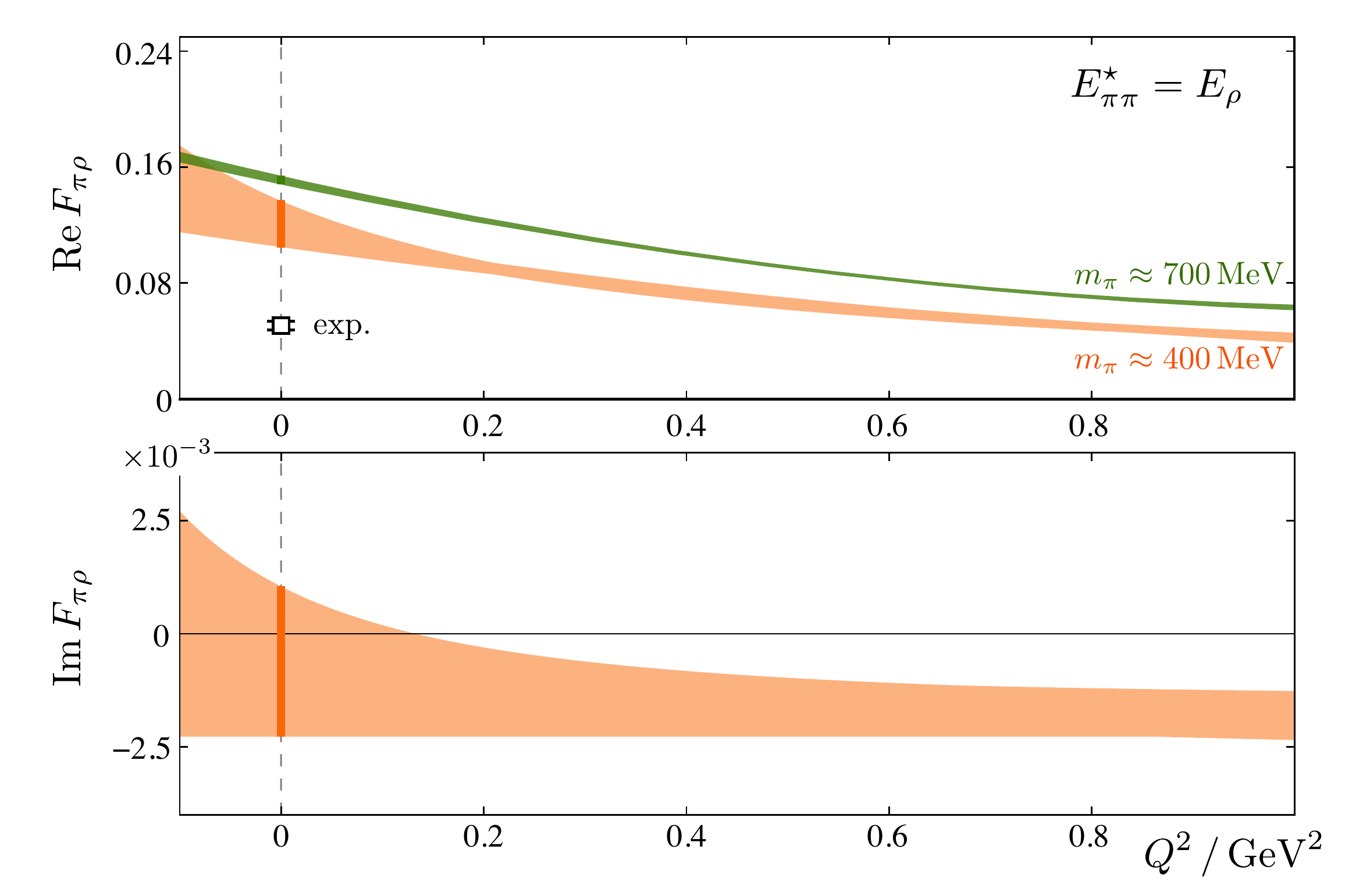}
\caption{The upper panel shows the real part of the form factor determined in this work [orange band] evaluated at the $\rho$ pole, $E_\rho/m_\pi=2.1805(32)-0.0151(5)i$. This is compared with the value obtained in Ref.~\cite{Shultz:2015pfa} [green band], where the $\rho$ resonance is QCD stable, and the experimentally determined $\rho\pi$ photocoupling~\cite{Huston:1986wi, Capraro:1987rp}. The lower panel shows the previously undetermined imaginary component of the form factor. }
\label{fig:Fpirho_Erho_BW}
\vspace*{-.6cm}
\end{center}
\end{figure}

With an analytic description of the $E^\star_{\pi\pi}$ dependence of form factor, we may analytically continue to the $\rho$ pole at $E^\star_{\pi\pi} = [ 2.1805(32)- i\,  0.0151(5) ] \, m_\pi$. The $Q^2$ dependence of the resulting form factor is shown in Fig.~\ref{fig:Fpirho_Erho_BW}, with the small imaginary part reflecting the fact that the $\rho$ resonance in this calculation is unstable, but with a small hadronic width -- as the pion mass is decreased in future lattice calculations~\cite{Wilson:2015dqa}, the width will increase and with it the imaginary part of the form factor.

\begin{figure}[t]
\begin{center}
\hspace*{-.7cm}                                                           
\includegraphics[scale=0.38]{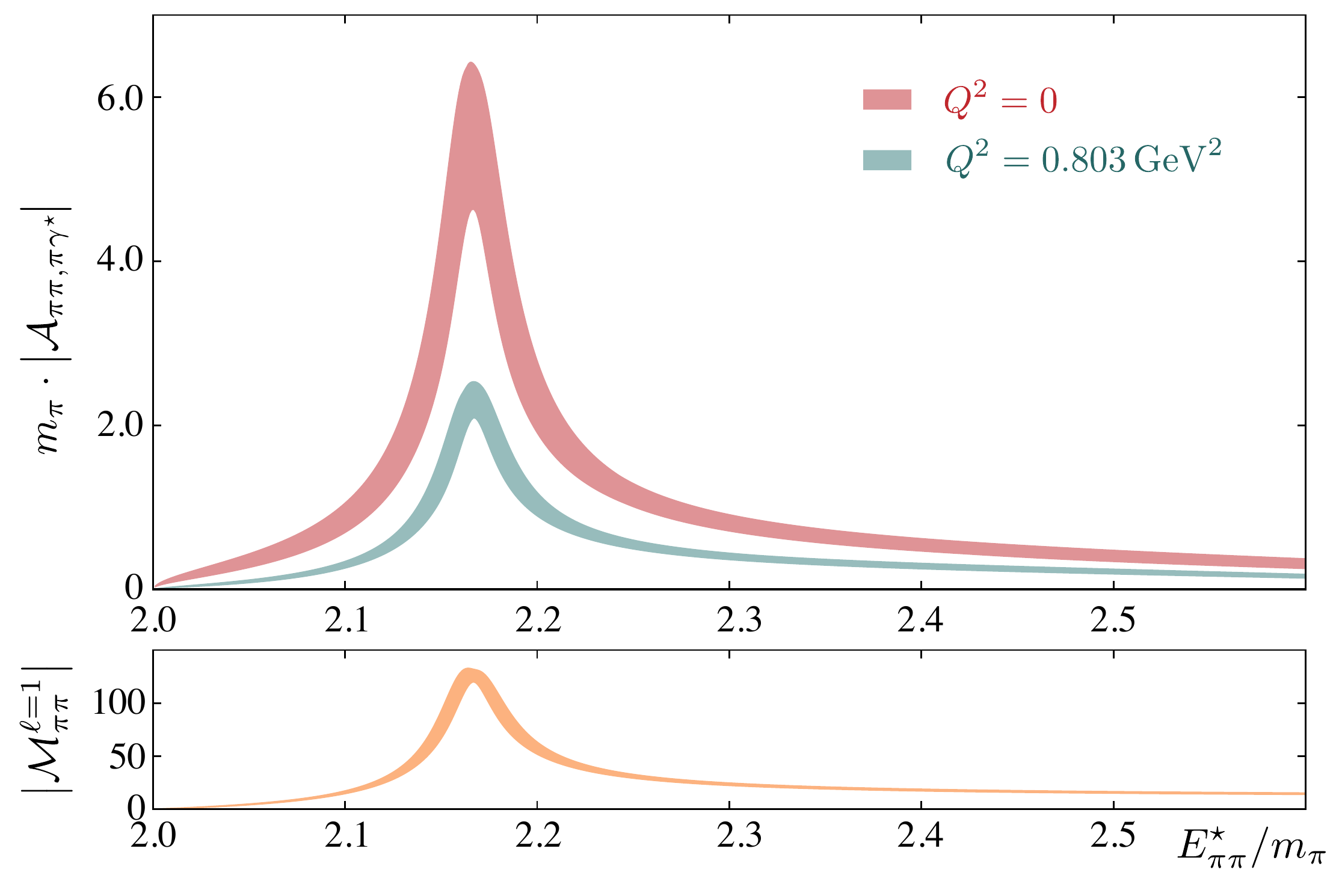}
\caption{ The top panel shows $ m_\pi  \big| {\mathcal{A}}_{\pi\pi,\pi\gamma^\star} \big|$ as a function of the c.m. $\pi\pi$ energy. This is determined for two different values of $Q^2/{\rm GeV}^2=0,0.803$. For comparison, in the lower panel we show the absolute values of the elastic $\ell=1$ $\pi\pi$ amplitude, $\big|\mathcal{M}_{\pi\pi}^{\ell=1}\big|$. }
\label{fig:slices_in_Q2}
\vspace*{-.6cm}
\end{center}
\end{figure}

The transition amplitude, ${\mathcal{A}}_{\pi\pi,\pi\gamma^\star}$, follows from Eq.~(\ref{eq:AtoF},) where the phase is fixed up to an overall sign by Watson's theorem to be the $\pi \pi \to \pi \pi$ phase shift. The remaining sign only has meaning in comparison to other transition amplitudes, and consequently, we need only present the absolute value of ${\mathcal{A}}_{\pi\pi,\pi\gamma^\star}$. In Fig.~\ref{fig:slices_in_Q2} we plot $ m_\pi  \big| {\mathcal{A}}_{\pi\pi,\pi\gamma^\star} \big|$ as a function of $E^\star_{\pi\pi}$ for two values of $Q^2$. This figure illustrates that as the $\pi\pi$ energy approaches the $\rho$ pole, the transition amplitude is dynamically enhanced by the resonance as one would expect. The resonant behavior, as a function of $E^\star_{\pi\pi}$, arises solely from the ${\RLL}$ factor in Eq.~(\ref{eq:Amp_to_matelem}); it is not due the parametrization in Eq.~(\ref{eq:AtoF}) which simply serves as the definition of the form factor

\begin{figure}[t]
\begin{center}
\hspace*{-.7cm}                                                           
\includegraphics[scale=0.38]{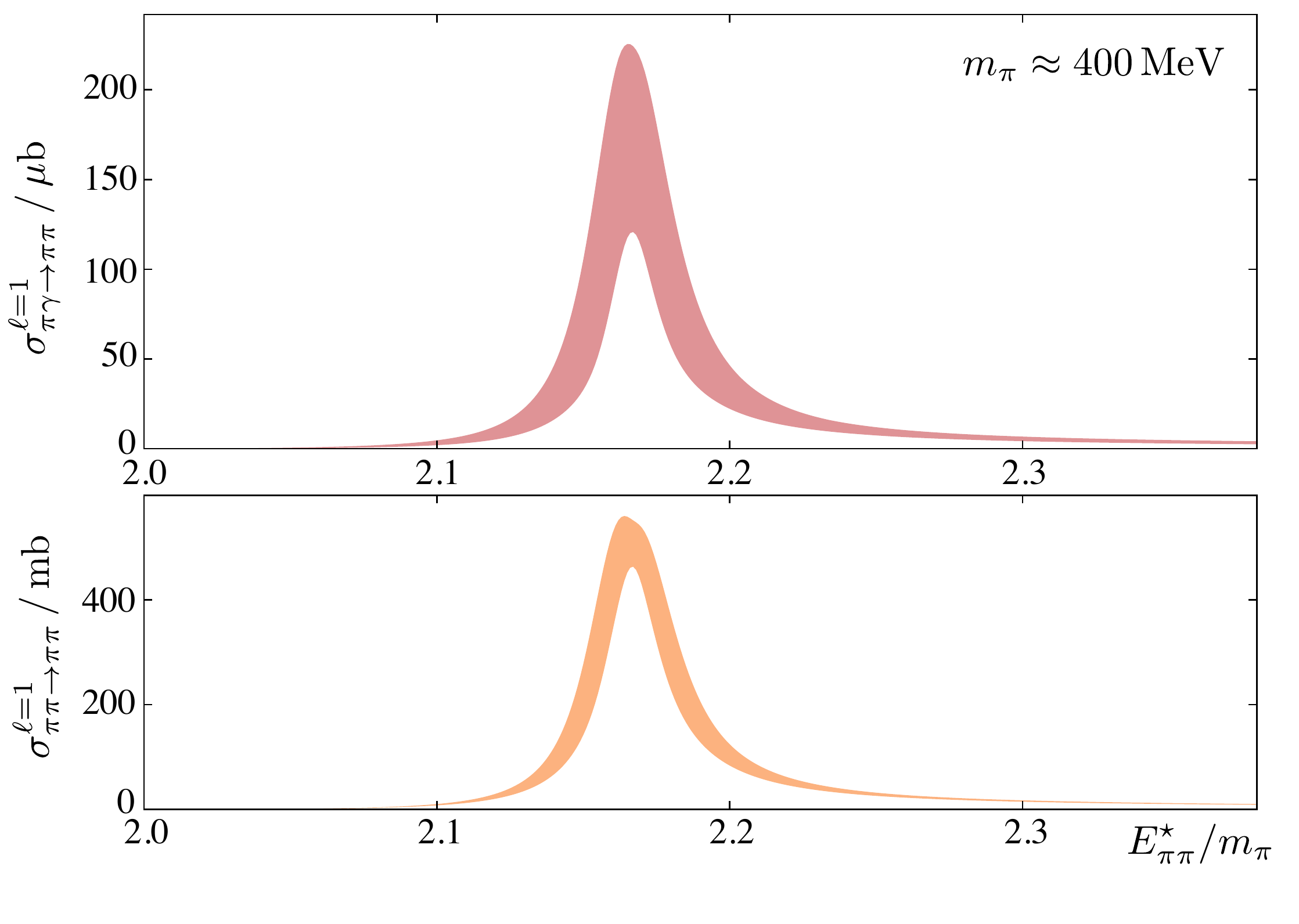}
\caption{ The top panel shows the $\pi^+\gamma\to\pi^+\pi^0$ cross section as a function of the $\pi\pi$ c.m. energy. The lower panel shows the elastic $\ell=1$ scattering cross section. One observes near the resonance the enhancement of the $\pi^+\gamma\to\pi^+\pi^0$ cross section. }\label{fig:cross_sec}
\end{center}
\vspace*{-.5cm}
\end{figure}

From Eq.~(\ref{eq:Apipipi}), one may readily obtain the $\pi^+\gamma\to\pi^+\pi^0$ cross section in terms of the reduced amplitude, ${\mathcal{A}}_{\pi\pi,\pi\gamma^\star}$ evaluated at  $Q^2=0$,
\begin{align}
\sigma(\pi^+\gamma\to\pi^+\pi^0) 
&= 
\alpha
\frac{{q}^\star_{\pi\pi}\,{q}^\star_{\pi\gamma}}{m_{\pi}^2}  
\Big| {\mathcal{A}}_{\pi\pi,\pi\gamma^\star}(E^{\star2}_{\pi\pi},0) \Big|^2,
 \label{eq:cross_section}
\end{align}
where ${q}^\star_{\pi\pi},\, {q}^\star_{\pi\gamma}$ are the c.m. relative momenta. In Fig.~\ref{fig:cross_sec} we plot this as a function of the c.m. energy. That the peak cross section for ${m_\pi \sim 400 \,\mathrm{MeV}}$ is significantly larger than phenomenological parametrizations of the physical cross section~\cite{Kaiser:2008ss, Hoferichter:2012pm} can be easily understood: near the resonance we have
\begin{align}
\lim_{E_{\pi\pi}^\star \to m_\rho}
\sigma(\pi^+\gamma\to\pi^+\pi^0) \propto  \frac{q^\star_{\pi\gamma} \, \F^{\,2}(m_{\rho},0)}{m_{\pi}^2~\Gamma_1(m_{\rho})}, \nonumber
\end{align}
and the $q^\star_{\pi\gamma} \, \F^2(m_{\rho},0)/m_{\pi}^2$ ratio we find to be approximately 60\% of the experimental value, and we expect this to vary only slowly with changing quark mass. On the other hand, the width of the $\rho$ resonance when $m_\pi \sim 400 \,\mathrm{MeV}$, $12.4(6)$~MeV~\cite{Dudek:2012gj}, is approximately 12 times smaller than the experimental width~\cite{pdg:2014}, scaling as expected for an approximately quark mass independent coupling, $g_{\rho \pi\pi}$, with reduced $P$-wave phase-space. This suggests that as future calculations are performed with quark masses closer to their physical values, and as the $\rho$ resonance becomes broader~\cite{Wilson:2015dqa}, the $\pi^+\gamma\to\pi^+\pi^0$ cross section will decrease by an order of magnitude. For comparison, in Fig.~\ref{fig:cross_sec} we plot the $\ell=1$ $\pi^+\pi^0$ elastic cross section, whose factor of 5 kinematic enhancement with respect to the experimental determination (see for example, Ref.~\cite{Protopopescu:1973sh}) can be understood by the $1/q^{\star2}$ dependence in the vicinity of the resonance.

 \paragraph{Final remarks:}
We have presented the first determination of a resonant radiative transition amplitude from QCD. This exploratory study of $\pi\pi\to\pi\gamma^\star$, although performed with unphysically heavy light quarks, serves as a proof of principle that hadronic transition processes involving resonating few-body states can be rigorously studied using lattice QCD. We have demonstrated how from this amplitude, by analytically continuing to a pole in the complex energy plane, one may obtain the $\rho\to\pi\gamma^\star$ form factor where the $\rho$ is treated as an unstable resonance, and have also obtained the $\pi^+\gamma \to \pi^+\pi^0$ cross section, and discussed how we expect the results to change in future calculations using lighter quark masses. 
 
Closely related techniques can be implemented in future studies of hadron structure and weak decays. As well as the obvious extension into the baryon sector, $\gamma^\star N \to N^\star \to N \pi$, there are processes important for testing the limits of the standard model such as ${B\rightarrow K\pi \, \ell^+\ell^-}$~\cite{Horgan:2013hoa,Horgan:2013pva}, where the $K\pi$ system is known to resonate.

Having demonstrated in this work the feasibility of studying radiative transition of two-body hadronic resonances directly from QCD, future studies will focus on the extension of this work. The technology for studying transition amplitudes with any number of open two-body states has been already developed~\cite{Briceno:2015csa, Briceno:2014uqa} and here we have tested it in the case where there is only one channel open. Future calculations will accommodate similar processes involving resonances that decay strongly to more than one hadronic state, for example $K\gamma^\star\to K^\star\to K\pi/K\eta$~\cite{Dudek:2014qha, Wilson:2014cna} and  $\pi\gamma^\star\to\rho^\star\to \pi\pi/K\overline{K}$~\cite{Wilson:2015dqa}. Furthermore, given the recent and exciting theoretical development for the study of three strongly interacting particles via lattice QCD~\cite{Polejaeva:2012ut,Briceno:2012rv,Hansen:2015zga,Hansen:2014eka}, we can also expect electromagnetic transition amplitudes involving three or more hadrons~(e.g.~$N\gamma^\star\to N^\star\to N\pi\pi$).~\footnote{We point the reader to a recent indirect calculation of $np\to d\gamma$ for heavy quark masses where both the isotriplet and isosinglet two-nucleon ground states are bound~\cite{Beane:2015yha}.}

\subsection*{Acknowledgments}

We thank our colleagues within the Hadron Spectrum Collaboration. 
The software codes
{\tt Chroma}~\cite{Edwards:2004sx} and {\tt QUDA}~\cite{Clark:2009wm,Babich:2010mu} were used to perform this work on clusters at Jefferson Laboratory under the USQCD Initiative and the LQCD ARRA project. We acknowledge resources used at the Oak Ridge Leadership Computing Facility, the National Center for Supercomputing Applications, the Texas Advanced Computer Center, and the Pittsburgh Supercomputer Center. 
Support is provided by U.S. Department of Energy Contract No. DE-AC05-06OR23177 under which Jefferson Science Associates manages Jefferson Lab, the Early Career award contract DE- SC0006765, and the JSA Graduate Fellowship program.  Support is also provided by the U.K. Science and Technology Facilities Council [Grant No. ST/L000385/1]. R. A. B. would like to thank M. Hansen, A. Rusetsky, S. Sharpe, W. Detmold,  I. V. Danilkin, Z. Davoudi, J. Goity, M. Pennington, and Stefan Meinel for useful discussions.


 
\bibliography{bibi}

\begin{thebibliography}{53}%
\makeatletter
\providecommand \@ifxundefined [1]{%
 \@ifx{#1\undefined}
}%
\providecommand \@ifnum [1]{%
 \ifnum #1\expandafter \@firstoftwo
 \else \expandafter \@secondoftwo
 \fi
}%
\providecommand \@ifx [1]{%
 \ifx #1\expandafter \@firstoftwo
 \else \expandafter \@secondoftwo
 \fi
}%
\providecommand \natexlab [1]{#1}%
\providecommand \enquote  [1]{``#1''}%
\providecommand \bibnamefont  [1]{#1}%
\providecommand \bibfnamefont [1]{#1}%
\providecommand \citenamefont [1]{#1}%
\providecommand \href@noop [0]{\@secondoftwo}%
\providecommand \href [0]{\begingroup \@sanitize@url \@href}%
\providecommand \@href[1]{\@@startlink{#1}\@@href}%
\providecommand \@@href[1]{\endgroup#1\@@endlink}%
\providecommand \@sanitize@url [0]{\catcode `\\12\catcode `\$12\catcode
  `\&12\catcode `\#12\catcode `\^12\catcode `\_12\catcode `\%12\relax}%
\providecommand \@@startlink[1]{}%
\providecommand \@@endlink[0]{}%
\providecommand \url  [0]{\begingroup\@sanitize@url \@url }%
\providecommand \@url [1]{\endgroup\@href {#1}{\urlprefix }}%
\providecommand \urlprefix  [0]{URL }%
\providecommand \Eprint [0]{\href }%
\providecommand \doibase [0]{http://dx.doi.org/}%
\providecommand \selectlanguage [0]{\@gobble}%
\providecommand \bibinfo  [0]{\@secondoftwo}%
\providecommand \bibfield  [0]{\@secondoftwo}%
\providecommand \translation [1]{[#1]}%
\providecommand \BibitemOpen [0]{}%
\providecommand \bibitemStop [0]{}%
\providecommand \bibitemNoStop [0]{.\EOS\space}%
\providecommand \EOS [0]{\spacefactor3000\relax}%
\providecommand \BibitemShut  [1]{\csname bibitem#1\endcsname}%
\let\auto@bib@innerbib\@empty
\bibitem [{\citenamefont {Aznauryan}\ \emph {et~al.}(2013)\citenamefont
  {Aznauryan}, \citenamefont {Bashir}, \citenamefont {Braun}, \citenamefont
  {Brodsky}, \citenamefont {Burkert} \emph {et~al.}}]{Aznauryan:2012ba}%
  \BibitemOpen
  \bibfield  {author} {\bibinfo {author} {\bibfnamefont {I.}~\bibnamefont
  {Aznauryan}}, \bibinfo {author} {\bibfnamefont {A.}~\bibnamefont {Bashir}},
  \bibinfo {author} {\bibfnamefont {V.}~\bibnamefont {Braun}}, \bibinfo
  {author} {\bibfnamefont {S.}~\bibnamefont {Brodsky}}, \bibinfo {author}
  {\bibfnamefont {V.}~\bibnamefont {Burkert}},  \emph {et~al.},\ }\href
  {\doibase 10.1142/S0218301313300154} {\bibfield  {journal} {\bibinfo
  {journal} {Int.J.Mod.Phys.}\ }\textbf {\bibinfo {volume} {E22}},\ \bibinfo
  {pages} {1330015} (\bibinfo {year} {2013})},\ \Eprint
  {http://arxiv.org/abs/1212.4891} {arXiv:1212.4891 [nucl-th]} \BibitemShut
  {NoStop}%
\bibitem [{\citenamefont {Colangelo}\ \emph
  {et~al.}(2014{\natexlab{a}})\citenamefont {Colangelo}, \citenamefont
  {Hoferichter}, \citenamefont {Procura},\ and\ \citenamefont
  {Stoffer}}]{Colangelo:2014dfa}%
  \BibitemOpen
  \bibfield  {author} {\bibinfo {author} {\bibfnamefont {G.}~\bibnamefont
  {Colangelo}}, \bibinfo {author} {\bibfnamefont {M.}~\bibnamefont
  {Hoferichter}}, \bibinfo {author} {\bibfnamefont {M.}~\bibnamefont
  {Procura}}, \ and\ \bibinfo {author} {\bibfnamefont {P.}~\bibnamefont
  {Stoffer}},\ }\href {\doibase 10.1007/JHEP09(2014)091} {\bibfield  {journal}
  {\bibinfo  {journal} {JHEP}\ }\textbf {\bibinfo {volume} {1409}},\ \bibinfo
  {pages} {091} (\bibinfo {year} {2014}{\natexlab{a}})},\ \Eprint
  {http://arxiv.org/abs/1402.7081} {arXiv:1402.7081 [hep-ph]} \BibitemShut
  {NoStop}%
\bibitem [{\citenamefont {Colangelo}\ \emph
  {et~al.}(2014{\natexlab{b}})\citenamefont {Colangelo}, \citenamefont
  {Hoferichter}, \citenamefont {Kubis}, \citenamefont {Procura},\ and\
  \citenamefont {Stoffer}}]{Colangelo:2014pva}%
  \BibitemOpen
  \bibfield  {author} {\bibinfo {author} {\bibfnamefont {G.}~\bibnamefont
  {Colangelo}}, \bibinfo {author} {\bibfnamefont {M.}~\bibnamefont
  {Hoferichter}}, \bibinfo {author} {\bibfnamefont {B.}~\bibnamefont {Kubis}},
  \bibinfo {author} {\bibfnamefont {M.}~\bibnamefont {Procura}}, \ and\
  \bibinfo {author} {\bibfnamefont {P.}~\bibnamefont {Stoffer}},\ }\href
  {\doibase 10.1016/j.physletb.2014.09.021} {\bibfield  {journal} {\bibinfo
  {journal} {Phys.Lett.}\ }\textbf {\bibinfo {volume} {B738}},\ \bibinfo
  {pages} {6} (\bibinfo {year} {2014}{\natexlab{b}})},\ \Eprint
  {http://arxiv.org/abs/1408.2517} {arXiv:1408.2517 [hep-ph]} \BibitemShut
  {NoStop}%
\bibitem [{\citenamefont {Wess}\ and\ \citenamefont
  {Zumino}(1971)}]{Wess:1971yu}%
  \BibitemOpen
  \bibfield  {author} {\bibinfo {author} {\bibfnamefont {J.}~\bibnamefont
  {Wess}}\ and\ \bibinfo {author} {\bibfnamefont {B.}~\bibnamefont {Zumino}},\
  }\href {\doibase 10.1016/0370-2693(71)90582-X} {\bibfield  {journal}
  {\bibinfo  {journal} {Phys.Lett.}\ }\textbf {\bibinfo {volume} {B37}},\
  \bibinfo {pages} {95} (\bibinfo {year} {1971})}\BibitemShut {NoStop}%
\bibitem [{\citenamefont {Witten}(1983)}]{Witten:1983tw}%
  \BibitemOpen
  \bibfield  {author} {\bibinfo {author} {\bibfnamefont {E.}~\bibnamefont
  {Witten}},\ }\href {\doibase 10.1016/0550-3213(83)90063-9} {\bibfield
  {journal} {\bibinfo  {journal} {Nucl.Phys.}\ }\textbf {\bibinfo {volume}
  {B223}},\ \bibinfo {pages} {422} (\bibinfo {year} {1983})}\BibitemShut
  {NoStop}%
\bibitem [{\citenamefont {Huston}\ \emph {et~al.}(1986)\citenamefont {Huston},
  \citenamefont {Berg}, \citenamefont {Chandlee}, \citenamefont {Cihangir},
  \citenamefont {Collick} \emph {et~al.}}]{Huston:1986wi}%
  \BibitemOpen
  \bibfield  {author} {\bibinfo {author} {\bibfnamefont {J.}~\bibnamefont
  {Huston}}, \bibinfo {author} {\bibfnamefont {D.}~\bibnamefont {Berg}},
  \bibinfo {author} {\bibfnamefont {C.}~\bibnamefont {Chandlee}}, \bibinfo
  {author} {\bibfnamefont {S.}~\bibnamefont {Cihangir}}, \bibinfo {author}
  {\bibfnamefont {B.}~\bibnamefont {Collick}},  \emph {et~al.},\ }\href
  {\doibase 10.1103/PhysRevD.33.3199} {\bibfield  {journal} {\bibinfo
  {journal} {Phys.Rev.}\ }\textbf {\bibinfo {volume} {D33}},\ \bibinfo {pages}
  {3199} (\bibinfo {year} {1986})}\BibitemShut {NoStop}%
\bibitem [{\citenamefont {Capraro}\ \emph {et~al.}(1987)\citenamefont
  {Capraro}, \citenamefont {Levy}, \citenamefont {Querrou}, \citenamefont
  {Van~Hecke}, \citenamefont {Verbeken} \emph {et~al.}}]{Capraro:1987rp}%
  \BibitemOpen
  \bibfield  {author} {\bibinfo {author} {\bibfnamefont {L.}~\bibnamefont
  {Capraro}}, \bibinfo {author} {\bibfnamefont {P.}~\bibnamefont {Levy}},
  \bibinfo {author} {\bibfnamefont {M.}~\bibnamefont {Querrou}}, \bibinfo
  {author} {\bibfnamefont {B.}~\bibnamefont {Van~Hecke}}, \bibinfo {author}
  {\bibfnamefont {M.}~\bibnamefont {Verbeken}},  \emph {et~al.},\ }\href
  {\doibase 10.1016/0550-3213(87)90233-1} {\bibfield  {journal} {\bibinfo
  {journal} {Nucl.Phys.}\ }\textbf {\bibinfo {volume} {B288}},\ \bibinfo
  {pages} {659} (\bibinfo {year} {1987})}\BibitemShut {NoStop}%
\bibitem [{\citenamefont {Marcucci}\ \emph {et~al.}(2015)\citenamefont
  {Marcucci}, \citenamefont {Gross}, \citenamefont {Pena}, \citenamefont
  {Piarulli}, \citenamefont {Schiavilla} \emph {et~al.}}]{Marcucci:2015rca}%
  \BibitemOpen
  \bibfield  {author} {\bibinfo {author} {\bibfnamefont {L.}~\bibnamefont
  {Marcucci}}, \bibinfo {author} {\bibfnamefont {F.}~\bibnamefont {Gross}},
  \bibinfo {author} {\bibfnamefont {M.}~\bibnamefont {Pena}}, \bibinfo {author}
  {\bibfnamefont {M.}~\bibnamefont {Piarulli}}, \bibinfo {author}
  {\bibfnamefont {R.}~\bibnamefont {Schiavilla}},  \emph {et~al.},\ }\href@noop
  {} {\  (\bibinfo {year} {2015})},\ \Eprint {http://arxiv.org/abs/1504.05063}
  {arXiv:1504.05063 [nucl-th]} \BibitemShut {NoStop}%
\bibitem [{\citenamefont {Shultz}\ \emph {et~al.}(2015)\citenamefont {Shultz},
  \citenamefont {Dudek},\ and\ \citenamefont {Edwards}}]{Shultz:2015pfa}%
  \BibitemOpen
  \bibfield  {author} {\bibinfo {author} {\bibfnamefont {C.~J.}\ \bibnamefont
  {Shultz}}, \bibinfo {author} {\bibfnamefont {J.~J.}\ \bibnamefont {Dudek}}, \
  and\ \bibinfo {author} {\bibfnamefont {R.~G.}\ \bibnamefont {Edwards}},\
  }\href {\doibase 10.1103/PhysRevD.91.114501} {\bibfield  {journal} {\bibinfo
  {journal} {Phys. Rev.}\ }\textbf {\bibinfo {volume} {D91}},\ \bibinfo {pages}
  {114501} (\bibinfo {year} {2015})},\ \Eprint
  {http://arxiv.org/abs/1501.07457} {arXiv:1501.07457 [hep-lat]} \BibitemShut
  {NoStop}%
\bibitem [{\citenamefont {Owen}\ \emph
  {et~al.}(2015{\natexlab{a}})\citenamefont {Owen}, \citenamefont {Kamleh},
  \citenamefont {Leinweber}, \citenamefont {Menadue},\ and\ \citenamefont
  {Mahbub}}]{Owen:2015gva}%
  \BibitemOpen
  \bibfield  {author} {\bibinfo {author} {\bibfnamefont {B.}~\bibnamefont
  {Owen}}, \bibinfo {author} {\bibfnamefont {W.}~\bibnamefont {Kamleh}},
  \bibinfo {author} {\bibfnamefont {D.}~\bibnamefont {Leinweber}}, \bibinfo
  {author} {\bibfnamefont {B.}~\bibnamefont {Menadue}}, \ and\ \bibinfo
  {author} {\bibfnamefont {S.}~\bibnamefont {Mahbub}},\ }\href {\doibase
  10.1103/PhysRevD.91.074503} {\bibfield  {journal} {\bibinfo  {journal}
  {Phys.Rev.}\ }\textbf {\bibinfo {volume} {D91}},\ \bibinfo {pages} {074503}
  (\bibinfo {year} {2015}{\natexlab{a}})},\ \Eprint
  {http://arxiv.org/abs/1501.02561} {arXiv:1501.02561 [hep-lat]} \BibitemShut
  {NoStop}%
\bibitem [{\citenamefont {Owen}\ \emph
  {et~al.}(2015{\natexlab{b}})\citenamefont {Owen}, \citenamefont {Kamleh},
  \citenamefont {Leinweber}, \citenamefont {Mahbub},\ and\ \citenamefont
  {Menadue}}]{Owen:2015fra}%
  \BibitemOpen
  \bibfield  {author} {\bibinfo {author} {\bibfnamefont {B.~J.}\ \bibnamefont
  {Owen}}, \bibinfo {author} {\bibfnamefont {W.}~\bibnamefont {Kamleh}},
  \bibinfo {author} {\bibfnamefont {D.~B.}\ \bibnamefont {Leinweber}}, \bibinfo
  {author} {\bibfnamefont {M.~S.}\ \bibnamefont {Mahbub}}, \ and\ \bibinfo
  {author} {\bibfnamefont {B.~J.}\ \bibnamefont {Menadue}},\ }\href {\doibase
  10.1103/PhysRevD.92.034513} {\bibfield  {journal} {\bibinfo  {journal} {Phys.
  Rev.}\ }\textbf {\bibinfo {volume} {D92}},\ \bibinfo {pages} {034513}
  (\bibinfo {year} {2015}{\natexlab{b}})},\ \Eprint
  {http://arxiv.org/abs/1505.02876} {arXiv:1505.02876 [hep-lat]} \BibitemShut
  {NoStop}%
\bibitem [{\citenamefont {L\"uscher}(1986)}]{Luscher:1986pf}%
  \BibitemOpen
  \bibfield  {author} {\bibinfo {author} {\bibfnamefont {M.}~\bibnamefont
  {L\"uscher}},\ }\href {\doibase 10.1007/BF01211097} {\bibfield  {journal}
  {\bibinfo  {journal} {Commun.Math.Phys.}\ }\textbf {\bibinfo {volume}
  {105}},\ \bibinfo {pages} {153} (\bibinfo {year} {1986})}\BibitemShut
  {NoStop}%
\bibitem [{\citenamefont {L\"uscher}(1991)}]{Luscher:1990ux}%
  \BibitemOpen
  \bibfield  {author} {\bibinfo {author} {\bibfnamefont {M.}~\bibnamefont
  {L\"uscher}},\ }\href {\doibase 10.1016/0550-3213(91)90366-6} {\bibfield
  {journal} {\bibinfo  {journal} {Nucl.Phys.}\ }\textbf {\bibinfo {volume}
  {B354}},\ \bibinfo {pages} {531} (\bibinfo {year} {1991})}\BibitemShut
  {NoStop}%
\bibitem [{\citenamefont {Rummukainen}\ and\ \citenamefont
  {Gottlieb}(1995)}]{Rummukainen:1995vs}%
  \BibitemOpen
  \bibfield  {author} {\bibinfo {author} {\bibfnamefont {K.}~\bibnamefont
  {Rummukainen}}\ and\ \bibinfo {author} {\bibfnamefont {S.~A.}\ \bibnamefont
  {Gottlieb}},\ }\href {\doibase 10.1016/0550-3213(95)00313-H} {\bibfield
  {journal} {\bibinfo  {journal} {Nucl. Phys.}\ }\textbf {\bibinfo {volume}
  {B450}},\ \bibinfo {pages} {397} (\bibinfo {year} {1995})},\ \Eprint
  {http://arxiv.org/abs/hep-lat/9503028} {arXiv:hep-lat/9503028} \BibitemShut
  {NoStop}%
\bibitem [{\citenamefont {Kim}\ \emph {et~al.}(2005)\citenamefont {Kim},
  \citenamefont {Sachrajda},\ and\ \citenamefont {Sharpe}}]{Kim:2005gf}%
  \BibitemOpen
  \bibfield  {author} {\bibinfo {author} {\bibfnamefont {C.}~\bibnamefont
  {Kim}}, \bibinfo {author} {\bibfnamefont {C.}~\bibnamefont {Sachrajda}}, \
  and\ \bibinfo {author} {\bibfnamefont {S.~R.}\ \bibnamefont {Sharpe}},\
  }\href {\doibase 10.1016/j.nuclphysb.2005.08.029} {\bibfield  {journal}
  {\bibinfo  {journal} {Nucl.Phys.}\ }\textbf {\bibinfo {volume} {B727}},\
  \bibinfo {pages} {218} (\bibinfo {year} {2005})},\ \Eprint
  {http://arxiv.org/abs/hep-lat/0507006} {arXiv:hep-lat/0507006 [hep-lat]}
  \BibitemShut {NoStop}%
\bibitem [{\citenamefont {Christ}\ \emph {et~al.}(2005)\citenamefont {Christ},
  \citenamefont {Kim},\ and\ \citenamefont {Yamazaki}}]{Christ:2005gi}%
  \BibitemOpen
  \bibfield  {author} {\bibinfo {author} {\bibfnamefont {N.~H.}\ \bibnamefont
  {Christ}}, \bibinfo {author} {\bibfnamefont {C.}~\bibnamefont {Kim}}, \ and\
  \bibinfo {author} {\bibfnamefont {T.}~\bibnamefont {Yamazaki}},\ }\href
  {\doibase 10.1103/PhysRevD.72.114506} {\bibfield  {journal} {\bibinfo
  {journal} {Phys.Rev.}\ }\textbf {\bibinfo {volume} {D72}},\ \bibinfo {pages}
  {114506} (\bibinfo {year} {2005})},\ \Eprint
  {http://arxiv.org/abs/hep-lat/0507009} {arXiv:hep-lat/0507009 [hep-lat]}
  \BibitemShut {NoStop}%
\bibitem [{\citenamefont {Brice\~no}\ and\ \citenamefont
  {Davoudi}(2013)}]{Briceno:2012yi}%
  \BibitemOpen
  \bibfield  {author} {\bibinfo {author} {\bibfnamefont {R.~A.}\ \bibnamefont
  {Brice\~no}}\ and\ \bibinfo {author} {\bibfnamefont {Z.}~\bibnamefont
  {Davoudi}},\ }\href {\doibase 10.1103/PhysRevD.88.094507} {\bibfield
  {journal} {\bibinfo  {journal} {Phys. Rev. D. 88,}\ }\textbf {\bibinfo
  {volume} {094507}},\ \bibinfo {pages} {094507} (\bibinfo {year} {2013})},\
  \Eprint {http://arxiv.org/abs/1204.1110} {arXiv:1204.1110 [hep-lat]}
  \BibitemShut {NoStop}%
\bibitem [{\citenamefont {Hansen}\ and\ \citenamefont
  {Sharpe}(2012)}]{Hansen:2012tf}%
  \BibitemOpen
  \bibfield  {author} {\bibinfo {author} {\bibfnamefont {M.~T.}\ \bibnamefont
  {Hansen}}\ and\ \bibinfo {author} {\bibfnamefont {S.~R.}\ \bibnamefont
  {Sharpe}},\ }\href {\doibase 10.1103/PhysRevD.86.016007} {\bibfield
  {journal} {\bibinfo  {journal} {Phys.Rev.}\ }\textbf {\bibinfo {volume}
  {D86}},\ \bibinfo {pages} {016007} (\bibinfo {year} {2012})},\ \Eprint
  {http://arxiv.org/abs/1204.0826} {arXiv:1204.0826 [hep-lat]} \BibitemShut
  {NoStop}%
\bibitem [{\citenamefont {Brice\~no}(2014)}]{Briceno:2014oea}%
  \BibitemOpen
  \bibfield  {author} {\bibinfo {author} {\bibfnamefont {R.~A.}\ \bibnamefont
  {Brice\~no}},\ }\href {\doibase 10.1103/PhysRevD.89.074507} {\bibfield
  {journal} {\bibinfo  {journal} {Phys.Rev.}\ }\textbf {\bibinfo {volume}
  {D89}},\ \bibinfo {pages} {074507} (\bibinfo {year} {2014})},\ \Eprint
  {http://arxiv.org/abs/1401.3312} {arXiv:1401.3312 [hep-lat]} \BibitemShut
  {NoStop}%
\bibitem [{\citenamefont {Polejaeva}\ and\ \citenamefont
  {Rusetsky}(2012)}]{Polejaeva:2012ut}%
  \BibitemOpen
  \bibfield  {author} {\bibinfo {author} {\bibfnamefont {K.}~\bibnamefont
  {Polejaeva}}\ and\ \bibinfo {author} {\bibfnamefont {A.}~\bibnamefont
  {Rusetsky}},\ }\href@noop {} {\bibfield  {journal} {\bibinfo  {journal}
  {Eur.Phys.J.}\ }\textbf {\bibinfo {volume} {A48}},\ \bibinfo {pages} {67}
  (\bibinfo {year} {2012})},\ \Eprint {http://arxiv.org/abs/1203.1241}
  {arXiv:1203.1241 [hep-lat]} \BibitemShut {NoStop}%
\bibitem [{\citenamefont {Brice\~no}\ and\ \citenamefont
  {Davoudi}(2012)}]{Briceno:2012rv}%
  \BibitemOpen
  \bibfield  {author} {\bibinfo {author} {\bibfnamefont {R.~A.}\ \bibnamefont
  {Brice\~no}}\ and\ \bibinfo {author} {\bibfnamefont {Z.}~\bibnamefont
  {Davoudi}},\ }\href {\doibase 10.1103/PhysRevD.87.094507} {\bibfield
  {journal} {\bibinfo  {journal} {Phys.Rev.}\ }\textbf {\bibinfo {volume}
  {D87}},\ \bibinfo {pages} {094507} (\bibinfo {year} {2012})},\ \Eprint
  {http://arxiv.org/abs/1212.3398} {arXiv:1212.3398 [hep-lat]} \BibitemShut
  {NoStop}%
\bibitem [{\citenamefont {Hansen}\ and\ \citenamefont
  {Sharpe}(2015)}]{Hansen:2015zga}%
  \BibitemOpen
  \bibfield  {author} {\bibinfo {author} {\bibfnamefont {M.~T.}\ \bibnamefont
  {Hansen}}\ and\ \bibinfo {author} {\bibfnamefont {S.~R.}\ \bibnamefont
  {Sharpe}},\ }\href@noop {} {\  (\bibinfo {year} {2015})},\ \Eprint
  {http://arxiv.org/abs/1504.04248} {arXiv:1504.04248 [hep-lat]} \BibitemShut
  {NoStop}%
\bibitem [{\citenamefont {Hansen}\ and\ \citenamefont
  {Sharpe}(2014)}]{Hansen:2014eka}%
  \BibitemOpen
  \bibfield  {author} {\bibinfo {author} {\bibfnamefont {M.~T.}\ \bibnamefont
  {Hansen}}\ and\ \bibinfo {author} {\bibfnamefont {S.~R.}\ \bibnamefont
  {Sharpe}},\ }\href {\doibase 10.1103/PhysRevD.90.116003} {\bibfield
  {journal} {\bibinfo  {journal} {Phys.Rev.}\ }\textbf {\bibinfo {volume}
  {D90}},\ \bibinfo {pages} {116003} (\bibinfo {year} {2014})},\ \Eprint
  {http://arxiv.org/abs/1408.5933} {arXiv:1408.5933 [hep-lat]} \BibitemShut
  {NoStop}%
\bibitem [{\citenamefont {Dudek}\ \emph {et~al.}(2014)\citenamefont {Dudek},
  \citenamefont {Edwards}, \citenamefont {Thomas},\ and\ \citenamefont
  {Wilson}}]{Dudek:2014qha}%
  \BibitemOpen
  \bibfield  {author} {\bibinfo {author} {\bibfnamefont {J.~J.}\ \bibnamefont
  {Dudek}}, \bibinfo {author} {\bibfnamefont {R.~G.}\ \bibnamefont {Edwards}},
  \bibinfo {author} {\bibfnamefont {C.~E.}\ \bibnamefont {Thomas}}, \ and\
  \bibinfo {author} {\bibfnamefont {D.~J.}\ \bibnamefont {Wilson}} (\bibinfo
  {collaboration} {Hadron Spectrum}),\ }\href {\doibase
  10.1103/PhysRevLett.113.182001} {\bibfield  {journal} {\bibinfo  {journal}
  {Phys. Rev. Lett.}\ }\textbf {\bibinfo {volume} {113}},\ \bibinfo {pages}
  {182001} (\bibinfo {year} {2014})},\ \Eprint {http://arxiv.org/abs/1406.4158}
  {arXiv:1406.4158 [hep-ph]} \BibitemShut {NoStop}%
\bibitem [{\citenamefont {Wilson}\ \emph
  {et~al.}(2015{\natexlab{a}})\citenamefont {Wilson}, \citenamefont {Dudek},
  \citenamefont {Edwards},\ and\ \citenamefont {Thomas}}]{Wilson:2014cna}%
  \BibitemOpen
  \bibfield  {author} {\bibinfo {author} {\bibfnamefont {D.~J.}\ \bibnamefont
  {Wilson}}, \bibinfo {author} {\bibfnamefont {J.~J.}\ \bibnamefont {Dudek}},
  \bibinfo {author} {\bibfnamefont {R.~G.}\ \bibnamefont {Edwards}}, \ and\
  \bibinfo {author} {\bibfnamefont {C.~E.}\ \bibnamefont {Thomas}},\ }\href
  {\doibase 10.1103/PhysRevD.91.054008} {\bibfield  {journal} {\bibinfo
  {journal} {Phys. Rev.}\ }\textbf {\bibinfo {volume} {D91}},\ \bibinfo {pages}
  {054008} (\bibinfo {year} {2015}{\natexlab{a}})},\ \Eprint
  {http://arxiv.org/abs/1411.2004} {arXiv:1411.2004 [hep-ph]} \BibitemShut
  {NoStop}%
\bibitem [{\citenamefont {Wilson}\ \emph
  {et~al.}(2015{\natexlab{b}})\citenamefont {Wilson}, \citenamefont {Briceno},
  \citenamefont {Dudek}, \citenamefont {Edwards},\ and\ \citenamefont
  {Thomas}}]{Wilson:2015dqa}%
  \BibitemOpen
  \bibfield  {author} {\bibinfo {author} {\bibfnamefont {D.~J.}\ \bibnamefont
  {Wilson}}, \bibinfo {author} {\bibfnamefont {R.~A.}\ \bibnamefont {Briceno}},
  \bibinfo {author} {\bibfnamefont {J.~J.}\ \bibnamefont {Dudek}}, \bibinfo
  {author} {\bibfnamefont {R.~G.}\ \bibnamefont {Edwards}}, \ and\ \bibinfo
  {author} {\bibfnamefont {C.~E.}\ \bibnamefont {Thomas}},\ }\href {\doibase
  10.1103/PhysRevD.92.094502} {\bibfield  {journal} {\bibinfo  {journal} {Phys.
  Rev.}\ }\textbf {\bibinfo {volume} {D92}},\ \bibinfo {pages} {094502}
  (\bibinfo {year} {2015}{\natexlab{b}})},\ \Eprint
  {http://arxiv.org/abs/1507.02599} {arXiv:1507.02599 [hep-ph]} \BibitemShut
  {NoStop}%
\bibitem [{\citenamefont {Lellouch}\ and\ \citenamefont
  {L\"uscher}(2001)}]{Lellouch:2000pv}%
  \BibitemOpen
  \bibfield  {author} {\bibinfo {author} {\bibfnamefont {L.}~\bibnamefont
  {Lellouch}}\ and\ \bibinfo {author} {\bibfnamefont {M.}~\bibnamefont
  {L\"uscher}},\ }\href@noop {} {\bibfield  {journal} {\bibinfo  {journal}
  {Commun.Math.Phys.}\ }\textbf {\bibinfo {volume} {219}},\ \bibinfo {pages}
  {31} (\bibinfo {year} {2001})},\ \Eprint
  {http://arxiv.org/abs/hep-lat/0003023} {arXiv:hep-lat/0003023 [hep-lat]}
  \BibitemShut {NoStop}%
\bibitem [{\citenamefont {Bai}\ \emph {et~al.}(2015)\citenamefont {Bai} \emph
  {et~al.}}]{Bai:2015nea}%
  \BibitemOpen
  \bibfield  {author} {\bibinfo {author} {\bibfnamefont {Z.}~\bibnamefont
  {Bai}} \emph {et~al.} (\bibinfo {collaboration} {RBC, UKQCD}),\ }\href
  {\doibase 10.1103/PhysRevLett.115.212001} {\bibfield  {journal} {\bibinfo
  {journal} {Phys. Rev. Lett.}\ }\textbf {\bibinfo {volume} {115}},\ \bibinfo
  {pages} {212001} (\bibinfo {year} {2015})},\ \Eprint
  {http://arxiv.org/abs/1505.07863} {arXiv:1505.07863 [hep-lat]} \BibitemShut
  {NoStop}%
\bibitem [{\citenamefont {Blum}\ \emph
  {et~al.}(2012{\natexlab{a}})\citenamefont {Blum}, \citenamefont {Boyle},
  \citenamefont {Christ}, \citenamefont {Garron}, \citenamefont {Goode} \emph
  {et~al.}}]{Blum:2012uk}%
  \BibitemOpen
  \bibfield  {author} {\bibinfo {author} {\bibfnamefont {T.}~\bibnamefont
  {Blum}}, \bibinfo {author} {\bibfnamefont {P.}~\bibnamefont {Boyle}},
  \bibinfo {author} {\bibfnamefont {N.}~\bibnamefont {Christ}}, \bibinfo
  {author} {\bibfnamefont {N.}~\bibnamefont {Garron}}, \bibinfo {author}
  {\bibfnamefont {E.}~\bibnamefont {Goode}},  \emph {et~al.},\ }\href {\doibase
  10.1103/PhysRevD.86.074513} {\bibfield  {journal} {\bibinfo  {journal}
  {Phys.Rev.}\ }\textbf {\bibinfo {volume} {D86}},\ \bibinfo {pages} {074513}
  (\bibinfo {year} {2012}{\natexlab{a}})},\ \Eprint
  {http://arxiv.org/abs/1206.5142} {arXiv:1206.5142 [hep-lat]} \BibitemShut
  {NoStop}%
\bibitem [{\citenamefont {Boyle}\ \emph {et~al.}(2013)\citenamefont {Boyle}
  \emph {et~al.}}]{Boyle:2012ys}%
  \BibitemOpen
  \bibfield  {author} {\bibinfo {author} {\bibfnamefont {P.}~\bibnamefont
  {Boyle}} \emph {et~al.} (\bibinfo {collaboration} {RBC, UKQCD}),\ }\href
  {\doibase 10.1103/PhysRevLett.110.152001} {\bibfield  {journal} {\bibinfo
  {journal} {Phys.Rev.Lett.}\ }\textbf {\bibinfo {volume} {110}},\ \bibinfo
  {pages} {152001} (\bibinfo {year} {2013})},\ \Eprint
  {http://arxiv.org/abs/1212.1474} {arXiv:1212.1474 [hep-lat]} \BibitemShut
  {NoStop}%
\bibitem [{\citenamefont {Blum}\ \emph {et~al.}(2011)\citenamefont {Blum},
  \citenamefont {Boyle}, \citenamefont {Christ}, \citenamefont {Garron},
  \citenamefont {Goode} \emph {et~al.}}]{Blum:2011pu}%
  \BibitemOpen
  \bibfield  {author} {\bibinfo {author} {\bibfnamefont {T.}~\bibnamefont
  {Blum}}, \bibinfo {author} {\bibfnamefont {P.}~\bibnamefont {Boyle}},
  \bibinfo {author} {\bibfnamefont {N.}~\bibnamefont {Christ}}, \bibinfo
  {author} {\bibfnamefont {N.}~\bibnamefont {Garron}}, \bibinfo {author}
  {\bibfnamefont {E.}~\bibnamefont {Goode}},  \emph {et~al.},\ }\href {\doibase
  10.1103/PhysRevD.84.114503} {\bibfield  {journal} {\bibinfo  {journal}
  {Phys.Rev.}\ }\textbf {\bibinfo {volume} {D84}},\ \bibinfo {pages} {114503}
  (\bibinfo {year} {2011})},\ \Eprint {http://arxiv.org/abs/1106.2714}
  {arXiv:1106.2714 [hep-lat]} \BibitemShut {NoStop}%
\bibitem [{\citenamefont {Blum}\ \emph
  {et~al.}(2012{\natexlab{b}})\citenamefont {Blum}, \citenamefont {Boyle},
  \citenamefont {Christ}, \citenamefont {Garron}, \citenamefont {Goode} \emph
  {et~al.}}]{Blum:2011ng}%
  \BibitemOpen
  \bibfield  {author} {\bibinfo {author} {\bibfnamefont {T.}~\bibnamefont
  {Blum}}, \bibinfo {author} {\bibfnamefont {P.}~\bibnamefont {Boyle}},
  \bibinfo {author} {\bibfnamefont {N.}~\bibnamefont {Christ}}, \bibinfo
  {author} {\bibfnamefont {N.}~\bibnamefont {Garron}}, \bibinfo {author}
  {\bibfnamefont {E.}~\bibnamefont {Goode}},  \emph {et~al.},\ }\href {\doibase
  10.1103/PhysRevLett.108.141601} {\bibfield  {journal} {\bibinfo  {journal}
  {Phys.Rev.Lett.}\ }\textbf {\bibinfo {volume} {108}},\ \bibinfo {pages}
  {141601} (\bibinfo {year} {2012}{\natexlab{b}})},\ \Eprint
  {http://arxiv.org/abs/1111.1699} {arXiv:1111.1699 [hep-lat]} \BibitemShut
  {NoStop}%
\bibitem [{\citenamefont {Lin}\ \emph {et~al.}(2001)\citenamefont {Lin},
  \citenamefont {Martinelli}, \citenamefont {Sachrajda},\ and\ \citenamefont
  {Testa}}]{Lin:2001ek}%
  \BibitemOpen
  \bibfield  {author} {\bibinfo {author} {\bibfnamefont {C.~D.}\ \bibnamefont
  {Lin}}, \bibinfo {author} {\bibfnamefont {G.}~\bibnamefont {Martinelli}},
  \bibinfo {author} {\bibfnamefont {C.~T.}\ \bibnamefont {Sachrajda}}, \ and\
  \bibinfo {author} {\bibfnamefont {M.}~\bibnamefont {Testa}},\ }\href
  {\doibase 10.1016/S0550-3213(01)00495-3} {\bibfield  {journal} {\bibinfo
  {journal} {Nucl.Phys.}\ }\textbf {\bibinfo {volume} {B619}},\ \bibinfo
  {pages} {467} (\bibinfo {year} {2001})},\ \Eprint
  {http://arxiv.org/abs/hep-lat/0104006} {arXiv:hep-lat/0104006 [hep-lat]}
  \BibitemShut {NoStop}%
\bibitem [{\citenamefont {Agadjanov}\ \emph {et~al.}(2014)\citenamefont
  {Agadjanov}, \citenamefont {Bernard}, \citenamefont {Mei$\ss$ner},\ and\
  \citenamefont {Rusetsky}}]{Agadjanov:2014kha}%
  \BibitemOpen
  \bibfield  {author} {\bibinfo {author} {\bibfnamefont {A.}~\bibnamefont
  {Agadjanov}}, \bibinfo {author} {\bibfnamefont {V.}~\bibnamefont {Bernard}},
  \bibinfo {author} {\bibfnamefont {U.-G.}\ \bibnamefont {Mei$\ss$ner}}, \ and\
  \bibinfo {author} {\bibfnamefont {A.}~\bibnamefont {Rusetsky}},\ }\href
  {\doibase 10.1016/j.nuclphysb.2014.07.023} {\bibfield  {journal} {\bibinfo
  {journal} {Nucl.Phys.}\ }\textbf {\bibinfo {volume} {B886}},\ \bibinfo
  {pages} {1199} (\bibinfo {year} {2014})},\ \Eprint
  {http://arxiv.org/abs/1405.3476} {arXiv:1405.3476 [hep-lat]} \BibitemShut
  {NoStop}%
\bibitem [{\citenamefont {Meyer}(2011)}]{Meyer:2011um}%
  \BibitemOpen
  \bibfield  {author} {\bibinfo {author} {\bibfnamefont {H.~B.}\ \bibnamefont
  {Meyer}},\ }\href {\doibase 10.1103/PhysRevLett.107.072002} {\bibfield
  {journal} {\bibinfo  {journal} {Phys. Rev. Lett.}\ }\textbf {\bibinfo
  {volume} {107}},\ \bibinfo {pages} {072002} (\bibinfo {year} {2011})},\
  \Eprint {http://arxiv.org/abs/1105.1892} {arXiv:1105.1892 [hep-lat]}
  \BibitemShut {NoStop}%
\bibitem [{\citenamefont {Meyer}(2012)}]{Meyer:2012wk}%
  \BibitemOpen
  \bibfield  {author} {\bibinfo {author} {\bibfnamefont {H.~B.}\ \bibnamefont
  {Meyer}},\ }\href@noop {} {\  (\bibinfo {year} {2012})},\ \Eprint
  {http://arxiv.org/abs/1202.6675} {arXiv:1202.6675 [hep-lat]} \BibitemShut
  {NoStop}%
\bibitem [{\citenamefont {Feng}\ \emph {et~al.}(2015)\citenamefont {Feng},
  \citenamefont {Aoki}, \citenamefont {Hashimoto},\ and\ \citenamefont
  {Kaneko}}]{Feng:2014gba}%
  \BibitemOpen
  \bibfield  {author} {\bibinfo {author} {\bibfnamefont {X.}~\bibnamefont
  {Feng}}, \bibinfo {author} {\bibfnamefont {S.}~\bibnamefont {Aoki}}, \bibinfo
  {author} {\bibfnamefont {S.}~\bibnamefont {Hashimoto}}, \ and\ \bibinfo
  {author} {\bibfnamefont {T.}~\bibnamefont {Kaneko}},\ }\href {\doibase
  10.1103/PhysRevD.91.054504} {\bibfield  {journal} {\bibinfo  {journal} {Phys.
  Rev.}\ }\textbf {\bibinfo {volume} {D91}},\ \bibinfo {pages} {054504}
  (\bibinfo {year} {2015})},\ \Eprint {http://arxiv.org/abs/1412.6319}
  {arXiv:1412.6319 [hep-lat]} \BibitemShut {NoStop}%
\bibitem [{\citenamefont {Briceño}\ and\ \citenamefont
  {Hansen}(2015)}]{Briceno:2015csa}%
  \BibitemOpen
  \bibfield  {author} {\bibinfo {author} {\bibfnamefont {R.~A.}\ \bibnamefont
  {Briceño}}\ and\ \bibinfo {author} {\bibfnamefont {M.~T.}\ \bibnamefont
  {Hansen}},\ }\href {\doibase 10.1103/PhysRevD.92.074509} {\bibfield
  {journal} {\bibinfo  {journal} {Phys. Rev.}\ }\textbf {\bibinfo {volume}
  {D92}},\ \bibinfo {pages} {074509} (\bibinfo {year} {2015})},\ \Eprint
  {http://arxiv.org/abs/1502.04314} {arXiv:1502.04314 [hep-lat]} \BibitemShut
  {NoStop}%
\bibitem [{\citenamefont {Brice\~no}\ \emph {et~al.}(2015)\citenamefont
  {Brice\~no}, \citenamefont {Hansen},\ and\ \citenamefont
  {Walker-Loud}}]{Briceno:2014uqa}%
  \BibitemOpen
  \bibfield  {author} {\bibinfo {author} {\bibfnamefont {R.~A.}\ \bibnamefont
  {Brice\~no}}, \bibinfo {author} {\bibfnamefont {M.~T.}\ \bibnamefont
  {Hansen}}, \ and\ \bibinfo {author} {\bibfnamefont {A.}~\bibnamefont
  {Walker-Loud}},\ }\href {\doibase 10.1103/PhysRevD.91.034501} {\bibfield
  {journal} {\bibinfo  {journal} {Phys.Rev.}\ }\textbf {\bibinfo {volume}
  {D91}},\ \bibinfo {pages} {034501} (\bibinfo {year} {2015})},\ \Eprint
  {http://arxiv.org/abs/1406.5965} {arXiv:1406.5965 [hep-lat]} \BibitemShut
  {NoStop}%
\bibitem [{\citenamefont {Dudek}\ \emph {et~al.}(2013)\citenamefont {Dudek},
  \citenamefont {Edwards},\ and\ \citenamefont {Thomas}}]{Dudek:2012xn}%
  \BibitemOpen
  \bibfield  {author} {\bibinfo {author} {\bibfnamefont {J.~J.}\ \bibnamefont
  {Dudek}}, \bibinfo {author} {\bibfnamefont {R.~G.}\ \bibnamefont {Edwards}},
  \ and\ \bibinfo {author} {\bibfnamefont {C.~E.}\ \bibnamefont {Thomas}},\
  }\href {\doibase 10.1103/PhysRevD.87.034505} {\bibfield  {journal} {\bibinfo
  {journal} {Phys.Rev.}\ }\textbf {\bibinfo {volume} {D87}},\ \bibinfo {pages}
  {034505} (\bibinfo {year} {2013})},\ \Eprint {http://arxiv.org/abs/1212.0830}
  {arXiv:1212.0830 [hep-ph]} \BibitemShut {NoStop}%
\bibitem [{\citenamefont {Lin}\ \emph {et~al.}(2009)\citenamefont {Lin} \emph
  {et~al.}}]{Lin:2008pr}%
  \BibitemOpen
  \bibfield  {author} {\bibinfo {author} {\bibfnamefont {H.-W.}\ \bibnamefont
  {Lin}} \emph {et~al.} (\bibinfo {collaboration} {Hadron Spectrum
  Collaboration}),\ }\href {\doibase 10.1103/PhysRevD.79.034502} {\bibfield
  {journal} {\bibinfo  {journal} {Phys.Rev.}\ }\textbf {\bibinfo {volume}
  {D79}},\ \bibinfo {pages} {034502} (\bibinfo {year} {2009})},\ \Eprint
  {http://arxiv.org/abs/0810.3588} {arXiv:0810.3588 [hep-lat]} \BibitemShut
  {NoStop}%
\bibitem [{\citenamefont {Dudek}\ \emph {et~al.}(2012)\citenamefont {Dudek},
  \citenamefont {Edwards},\ and\ \citenamefont {Thomas}}]{Dudek:2012gj}%
  \BibitemOpen
  \bibfield  {author} {\bibinfo {author} {\bibfnamefont {J.~J.}\ \bibnamefont
  {Dudek}}, \bibinfo {author} {\bibfnamefont {R.~G.}\ \bibnamefont {Edwards}},
  \ and\ \bibinfo {author} {\bibfnamefont {C.~E.}\ \bibnamefont {Thomas}},\
  }\href {\doibase 10.1103/PhysRevD.86.034031} {\bibfield  {journal} {\bibinfo
  {journal} {Phys.Rev.}\ }\textbf {\bibinfo {volume} {D86}},\ \bibinfo {pages}
  {034031} (\bibinfo {year} {2012})},\ \Eprint {http://arxiv.org/abs/1203.6041}
  {arXiv:1203.6041 [hep-ph]} \BibitemShut {NoStop}%
\bibitem [{\citenamefont {Thomas}\ \emph {et~al.}(2012)\citenamefont {Thomas},
  \citenamefont {Edwards},\ and\ \citenamefont {Dudek}}]{Thomas:2011rh}%
  \BibitemOpen
  \bibfield  {author} {\bibinfo {author} {\bibfnamefont {C.~E.}\ \bibnamefont
  {Thomas}}, \bibinfo {author} {\bibfnamefont {R.~G.}\ \bibnamefont {Edwards}},
  \ and\ \bibinfo {author} {\bibfnamefont {J.~J.}\ \bibnamefont {Dudek}},\
  }\href {\doibase 10.1103/PhysRevD.85.014507, 10.1103/PhysRevD.85.039901}
  {\bibfield  {journal} {\bibinfo  {journal} {Phys.Rev.}\ }\textbf {\bibinfo
  {volume} {D85}},\ \bibinfo {pages} {014507} (\bibinfo {year} {2012})},\
  \Eprint {http://arxiv.org/abs/1107.1930} {arXiv:1107.1930 [hep-lat]}
  \BibitemShut {NoStop}%
\bibitem [{\citenamefont {Kaiser}\ and\ \citenamefont
  {Friedrich}(2008)}]{Kaiser:2008ss}%
  \BibitemOpen
  \bibfield  {author} {\bibinfo {author} {\bibfnamefont {N.}~\bibnamefont
  {Kaiser}}\ and\ \bibinfo {author} {\bibfnamefont {J.}~\bibnamefont
  {Friedrich}},\ }\href {\doibase 10.1140/epja/i2008-10582-9} {\bibfield
  {journal} {\bibinfo  {journal} {Eur.Phys.J.}\ }\textbf {\bibinfo {volume}
  {A36}},\ \bibinfo {pages} {181} (\bibinfo {year} {2008})},\ \Eprint
  {http://arxiv.org/abs/0803.0995} {arXiv:0803.0995 [nucl-th]} \BibitemShut
  {NoStop}%
\bibitem [{\citenamefont {Hoferichter}\ \emph {et~al.}(2012)\citenamefont
  {Hoferichter}, \citenamefont {Kubis},\ and\ \citenamefont
  {Sakkas}}]{Hoferichter:2012pm}%
  \BibitemOpen
  \bibfield  {author} {\bibinfo {author} {\bibfnamefont {M.}~\bibnamefont
  {Hoferichter}}, \bibinfo {author} {\bibfnamefont {B.}~\bibnamefont {Kubis}},
  \ and\ \bibinfo {author} {\bibfnamefont {D.}~\bibnamefont {Sakkas}},\ }\href
  {\doibase 10.1103/PhysRevD.86.116009} {\bibfield  {journal} {\bibinfo
  {journal} {Phys.Rev.}\ }\textbf {\bibinfo {volume} {D86}},\ \bibinfo {pages}
  {116009} (\bibinfo {year} {2012})},\ \Eprint {http://arxiv.org/abs/1210.6793}
  {arXiv:1210.6793 [hep-ph]} \BibitemShut {NoStop}%
\bibitem [{\citenamefont {Olive}\ and\ \citenamefont {Group}(2014)}]{pdg:2014}%
  \BibitemOpen
  \bibfield  {author} {\bibinfo {author} {\bibfnamefont {K.}~\bibnamefont
  {Olive}}\ and\ \bibinfo {author} {\bibfnamefont {P.~D.}\ \bibnamefont
  {Group}},\ }\href {http://stacks.iop.org/1674-1137/38/i=9/a=090001}
  {\bibfield  {journal} {\bibinfo  {journal} {Chinese Physics C}\ }\textbf
  {\bibinfo {volume} {38}},\ \bibinfo {pages} {090001} (\bibinfo {year}
  {2014})}\BibitemShut {NoStop}%
\bibitem [{\citenamefont {Protopopescu}\ \emph {et~al.}(1973)\citenamefont
  {Protopopescu}, \citenamefont {Alston-Garnjost}, \citenamefont
  {Barbaro-Galtieri}, \citenamefont {Flatte}, \citenamefont {Friedman} \emph
  {et~al.}}]{Protopopescu:1973sh}%
  \BibitemOpen
  \bibfield  {author} {\bibinfo {author} {\bibfnamefont {S.}~\bibnamefont
  {Protopopescu}}, \bibinfo {author} {\bibfnamefont {M.}~\bibnamefont
  {Alston-Garnjost}}, \bibinfo {author} {\bibfnamefont {A.}~\bibnamefont
  {Barbaro-Galtieri}}, \bibinfo {author} {\bibfnamefont {S.~M.}\ \bibnamefont
  {Flatte}}, \bibinfo {author} {\bibfnamefont {J.}~\bibnamefont {Friedman}},
  \emph {et~al.},\ }\href {\doibase 10.1103/PhysRevD.7.1279} {\bibfield
  {journal} {\bibinfo  {journal} {Phys.Rev.}\ }\textbf {\bibinfo {volume}
  {D7}},\ \bibinfo {pages} {1279} (\bibinfo {year} {1973})}\BibitemShut
  {NoStop}%
\bibitem [{\citenamefont {Horgan}\ \emph
  {et~al.}(2014{\natexlab{a}})\citenamefont {Horgan}, \citenamefont {Liu},
  \citenamefont {Meinel},\ and\ \citenamefont {Wingate}}]{Horgan:2013hoa}%
  \BibitemOpen
  \bibfield  {author} {\bibinfo {author} {\bibfnamefont {R.~R.}\ \bibnamefont
  {Horgan}}, \bibinfo {author} {\bibfnamefont {Z.}~\bibnamefont {Liu}},
  \bibinfo {author} {\bibfnamefont {S.}~\bibnamefont {Meinel}}, \ and\ \bibinfo
  {author} {\bibfnamefont {M.}~\bibnamefont {Wingate}},\ }\href {\doibase
  10.1103/PhysRevD.89.094501} {\bibfield  {journal} {\bibinfo  {journal}
  {Phys.Rev.}\ }\textbf {\bibinfo {volume} {D89}},\ \bibinfo {pages} {094501}
  (\bibinfo {year} {2014}{\natexlab{a}})},\ \Eprint
  {http://arxiv.org/abs/1310.3722} {arXiv:1310.3722 [hep-lat]} \BibitemShut
  {NoStop}%
\bibitem [{\citenamefont {Horgan}\ \emph
  {et~al.}(2014{\natexlab{b}})\citenamefont {Horgan}, \citenamefont {Liu},
  \citenamefont {Meinel},\ and\ \citenamefont {Wingate}}]{Horgan:2013pva}%
  \BibitemOpen
  \bibfield  {author} {\bibinfo {author} {\bibfnamefont {R.~R.}\ \bibnamefont
  {Horgan}}, \bibinfo {author} {\bibfnamefont {Z.}~\bibnamefont {Liu}},
  \bibinfo {author} {\bibfnamefont {S.}~\bibnamefont {Meinel}}, \ and\ \bibinfo
  {author} {\bibfnamefont {M.}~\bibnamefont {Wingate}},\ }\href {\doibase
  10.1103/PhysRevLett.112.212003} {\bibfield  {journal} {\bibinfo  {journal}
  {Phys.Rev.Lett.}\ }\textbf {\bibinfo {volume} {112}},\ \bibinfo {pages}
  {212003} (\bibinfo {year} {2014}{\natexlab{b}})},\ \Eprint
  {http://arxiv.org/abs/1310.3887} {arXiv:1310.3887 [hep-ph]} \BibitemShut
  {NoStop}%
\bibitem [{\citenamefont {Beane}\ \emph {et~al.}(2015)\citenamefont {Beane},
  \citenamefont {Chang}, \citenamefont {Detmold}, \citenamefont {Orginos},
  \citenamefont {Parreño}, \citenamefont {Savage},\ and\ \citenamefont
  {Tiburzi}}]{Beane:2015yha}%
  \BibitemOpen
  \bibfield  {author} {\bibinfo {author} {\bibfnamefont {S.~R.}\ \bibnamefont
  {Beane}}, \bibinfo {author} {\bibfnamefont {E.}~\bibnamefont {Chang}},
  \bibinfo {author} {\bibfnamefont {W.}~\bibnamefont {Detmold}}, \bibinfo
  {author} {\bibfnamefont {K.}~\bibnamefont {Orginos}}, \bibinfo {author}
  {\bibfnamefont {A.}~\bibnamefont {Parreño}}, \bibinfo {author} {\bibfnamefont
  {M.~J.}\ \bibnamefont {Savage}}, \ and\ \bibinfo {author} {\bibfnamefont
  {B.~C.}\ \bibnamefont {Tiburzi}} (\bibinfo {collaboration} {NPLQCD}),\ }\href
  {\doibase 10.1103/PhysRevLett.115.132001} {\bibfield  {journal} {\bibinfo
  {journal} {Phys. Rev. Lett.}\ }\textbf {\bibinfo {volume} {115}},\ \bibinfo
  {pages} {132001} (\bibinfo {year} {2015})},\ \Eprint
  {http://arxiv.org/abs/1505.02422} {arXiv:1505.02422 [hep-lat]} \BibitemShut
  {NoStop}%
\bibitem [{\citenamefont {Edwards}\ and\ \citenamefont
  {Joo}(2005)}]{Edwards:2004sx}%
  \BibitemOpen
  \bibfield  {author} {\bibinfo {author} {\bibfnamefont {R.~G.}\ \bibnamefont
  {Edwards}}\ and\ \bibinfo {author} {\bibfnamefont {B.}~\bibnamefont {Joo}}
  (\bibinfo {collaboration} {SciDAC, LHPC, UKQCD}),\ }\bibfield  {booktitle}
  {\emph {\bibinfo {booktitle} {{Lattice field theory. Proceedings, 22nd
  International Symposium, Lattice 2004, Batavia, USA, June 21-26, 2004}}},\
  }\href {\doibase 10.1016/j.nuclphysbps.2004.11.254} {\bibfield  {journal}
  {\bibinfo  {journal} {Nucl. Phys. Proc. Suppl.}\ }\textbf {\bibinfo {volume}
  {140}},\ \bibinfo {pages} {832} (\bibinfo {year} {2005})},\ \bibinfo {note}
  {[,832(2004)]},\ \Eprint {http://arxiv.org/abs/hep-lat/0409003}
  {arXiv:hep-lat/0409003 [hep-lat]} \BibitemShut {NoStop}%
\bibitem [{\citenamefont {Clark}\ \emph {et~al.}(2010)\citenamefont {Clark},
  \citenamefont {Babich}, \citenamefont {Barros}, \citenamefont {Brower},\ and\
  \citenamefont {Rebbi}}]{Clark:2009wm}%
  \BibitemOpen
  \bibfield  {author} {\bibinfo {author} {\bibfnamefont {M.~A.}\ \bibnamefont
  {Clark}}, \bibinfo {author} {\bibfnamefont {R.}~\bibnamefont {Babich}},
  \bibinfo {author} {\bibfnamefont {K.}~\bibnamefont {Barros}}, \bibinfo
  {author} {\bibfnamefont {R.~C.}\ \bibnamefont {Brower}}, \ and\ \bibinfo
  {author} {\bibfnamefont {C.}~\bibnamefont {Rebbi}},\ }\href {\doibase
  10.1016/j.cpc.2010.05.002} {\bibfield  {journal} {\bibinfo  {journal}
  {Comput. Phys. Commun.}\ }\textbf {\bibinfo {volume} {181}},\ \bibinfo
  {pages} {1517} (\bibinfo {year} {2010})},\ \Eprint
  {http://arxiv.org/abs/0911.3191} {arXiv:0911.3191 [hep-lat]} \BibitemShut
  {NoStop}%
\bibitem [{\citenamefont {Babich}\ \emph {et~al.}(2010)\citenamefont {Babich},
  \citenamefont {Clark},\ and\ \citenamefont {Joo}}]{Babich:2010mu}%
  \BibitemOpen
  \bibfield  {author} {\bibinfo {author} {\bibfnamefont {R.}~\bibnamefont
  {Babich}}, \bibinfo {author} {\bibfnamefont {M.~A.}\ \bibnamefont {Clark}}, \
  and\ \bibinfo {author} {\bibfnamefont {B.}~\bibnamefont {Joo}},\ }in\ \href
  {http://dx.doi.org/10.1109/SC.2010.40} {\emph {\bibinfo {booktitle} {{SC 10
  (Supercomputing 2010) New Orleans, Louisiana, November 13-19, 2010}}}}\
  (\bibinfo {year} {2010})\ \Eprint {http://arxiv.org/abs/1011.0024}
  {arXiv:1011.0024 [hep-lat]} \BibitemShut {NoStop}%
\end{thebibliography}%

\end{document}